\documentclass[conference]{IEEEtran}
\IEEEoverridecommandlockouts
\usepackage{cite}
\usepackage{amsmath,amssymb,amsfonts}
\usepackage{algorithmic}
\usepackage{graphicx}
\usepackage{textcomp}
\usepackage{xcolor}
\usepackage{bm}
\usepackage{adjustbox}
\usepackage{algorithm}
\def\BibTeX{{\rm B\kern-.05em{\sc i\kern-.025em b}\kern-.08em
    T\kern-.1667em\lower.7ex\hbox{E}\kern-.125emX}}

\begin{document}

\title{
Pulse Shape-Aided Multipath Delay Estimation for Fine-Grained WiFi Sensing
}

\author{Ke Xu\textsuperscript{\textasteriskcentered}, He (Henry) Chen\textsuperscript{\textasteriskcentered\textdagger}, Chenshu Wu\textsuperscript{\textdaggerdbl}\\
\textsuperscript{\textasteriskcentered}Department of Information Engineering, The Chinese University of Hong Kong\\
\textsuperscript{\textdagger}Shun Hing Institute of Advanced Engineering, The Chinese University of Hong Kong\\
\textsuperscript{\textdaggerdbl}Department of Computer Science, The University of Hong Kong\\
\{xk020, he.chen\}@ie.cuhk.edu.hk, chenshu@cs.hku.hk\thanks{This research was supported in part by project \#MMT 79/22 of the Shun Hing Institute of Advanced Engineering, The Chinese University of Hong Kong. The authors would like to thank Soung Chang Liew for his insightful discussions on channel sparsity and the impacts of pulse shaping.}
}

\maketitle

\begin{abstract}
Due to the finite bandwidth of practical wireless systems, one multipath component can manifest itself as a discrete pulse consisting of multiple taps in the digital delay domain. This effect is called \textit{channel leakage}, which complicates the multipath delay estimation problem. In this paper, we develop a new algorithm to estimate multipath delays of leaked channels by leveraging the knowledge of pulse-shaping functions, which can be used to support fine-grained WiFi sensing applications. Specifically, we express the channel impulse response (CIR) as a linear combination of overcomplete basis vectors corresponding to different delays. Considering the limited number of paths in physical environments, we formulate the multipath delay estimation as a sparse recovery problem.
We then propose a sparse Bayesian learning (SBL) method to estimate the sparse vector and determine the number of physical paths and their associated delay parameters from the positions of the nonzero entries in the sparse vector. 
Simulation results show that our algorithm can accurately determine the number of paths, and achieve superior accuracy in path delay estimation and channel reconstruction compared to two benchmarking schemes.

\end{abstract}

\begin{IEEEkeywords}
channel leakage, multipath delay estimation, pulse shaping, sparse recovery
\end{IEEEkeywords}

\section{Introduction}

In the past years, WiFi has evolved beyond its initial role of providing connectivity among wireless devices to also encompass the capability of sensing surrounding environments \cite{ma2019wifi,tan2022commodity}. This new trend has facilitated various applications such as indoor localization, human gesture recognition, and vital sign detection, making WiFi a key enabling technology in the era of the Internet of things. 
In current WiFi systems, orthogonal frequency division multiplexing (OFDM) is used to combat frequency-selective fading. The channel frequency response in WiFi OFDM systems is often referred to as channel state information (CSI) in the sensing area.
Recently, several tools have been developed to extract the CSI from commodity WiFi devices\cite{halperin2011tool,xie2015precise}. These complex-valued CSI can provide fine-grained information of the environment, and has been widely used in WiFi sensing.

In the field of indoor localization, the path delay is a key parameter to determine the position of a target because it reflects the distance between the target and a WiFi device. 
However, it is nontrivial to obtain an accurate estimate of the path delays from CSI. In wireless systems, pulse shaping and matched filtering are performed at the transmitter and the receiver, respectively. Due to the limited system bandwidth, when the delay of a physical path is a non-integer multiple of the sampling period, the multipath component in the discrete delay domain will manifest itself as a pulse consisting of multiple taps, instead of a single tap. This effect is called \textit{channel leakage}\cite{taubock2010compressive}. 
\begin{figure}
\centering
\includegraphics[width=0.8\columnwidth]{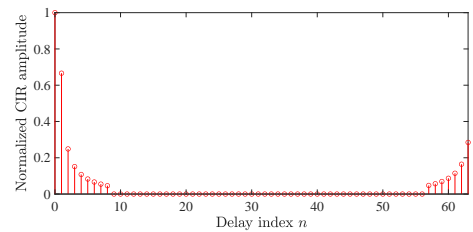}
\caption{Illustration of the channel leakage effect with a path delay of $20~\mathrm{ns}$ and a sampling period of $50~\mathrm{ns}$. A truncated raised-cosine filter with a roll-off factor of 0.05 and a length of 16 is used. In WiFi systems, packet detection is performed by a cross-correlation operation to identify the instance with the greatest signal strength. Therefore, the first tap corresponds to the sample with the largest amplitude in the pulse, and the taps preceding this point are shifted to the end of the CIR. As a result, the absolute delay information is lost due to the packet detection.}
\label{leakage}
\vspace{-2em}
\end{figure}
An example of the channel impulse response (CIR) with the leakage effect is illustrated in Fig. \ref{leakage}, in which the system has a sampling period of $T=50~\mathrm{ns}$ while a path arrives at $\tau=20~\mathrm{ns}$. Consequently, a pulse consisting of 16 taps is produced. 

Based on the above observation, a significant problem arises if two taps within the same pulse are recognized as two physical paths with distinct delays. This situation can lead to severe degradation of the localization accuracy, especially in multipath-aided applications such as \cite{zhang2022toward}.  Although existing subspace-based methods \cite{li2022superresolution} can directly estimate the delay parameters from the frequency domain and circumvent the above issue, these algorithms rely on the underlying assumption of using an ideal pulse shaping filter with a flat frequency response on data subcarriers. 
However, practical pulse functions have a finite time duration, incurring ripples in the passband and thus breaking the fundamental assumptions of those subspace-based methods. In \cite{li2021improved}, the authors developed an atomic norm-based approach to estimate the channel and further obtain the multipath parameters by incorporating the effect of pulse shaping. However, their algorithm focuses on single-carrier systems and also imposes stringent requirements on the pulse parameters of the widely-used raised-cosine filter, largely limiting their practical applications.

In this paper, we leverage the knowledge of the pulse shape and propose a new algorithm for multipath delay estimation. Since the CIR can be regarded as a superposition of multiple pulses shifted by different delays, we discretize the delay parameter into a set of grid points and formulate 
a sparse recovery problem using overcomplete basis vectors composed of discretized pulses shifted by the delays in the grid. We apply the sparse Bayesian learning (SBL) method \cite{tipping2001sparse} to update the weights of the basis vectors. The delay parameters of the physical paths can be automatically identified from the positions of nonzero weights, without requiring additional complex algorithms (e.g., \cite{stoica2004model})  to determine the number of~paths.

It is worth noting that the effect of pulse shaping has been investigated in conventional sparse channel estimation works (e.g.,  \cite{taubock2010compressive,scaglione2011compressed}). Indeed, the orthogonal matching pursuit (OMP)-based method proposed in \cite{scaglione2011compressed} can be adapted to address the problem in this paper. Nevertheless, as shown in Section \ref{simulation}, the method achieves inferior performance for delay estimation due to its greedy nature, even when the number of paths is given.

\section{Signal Model}\label{sigmal model}

We consider a physical multipath channel in the delay domain
\begin{equation}\label{physical channel}
    h_p(\tau)=\sum_{\ell=0}^{L-1}\alpha_\ell\delta(\tau-\tau_\ell),
\end{equation}
where $L$ is the number of paths, while $\alpha_\ell$ and $\tau_\ell$ are the complex amplitude and the delay of the $\ell$-th path, respectively. A composite channel incorporating the effect of pulse shaping can be expressed as
\begin{equation}\label{composite channel}
\begin{aligned}
    h(\tau)&=h_p(\tau)\otimes g_t(\tau)\otimes g_r(\tau)= h_p(\tau)\otimes g(\tau)\\
    &=\sum_{\ell=0}^{L-1}\alpha_\ell g(\tau-\tau_\ell),
\end{aligned}
\end{equation}
where $\otimes$ denotes the convolution operation, $g_t(\tau)$ and $g_r(\tau)$ are the pulse shaping filter at the transmitter and the matched filter at the receiver, respectively, and $g(\tau)\triangleq g_t(\tau)\otimes g_r(\tau)$. In this paper, we consider a truncated raised-cosine pulse function, i.e.,
$g(t)=\mathrm{sinc}\left(\frac{t}{T}\right)\frac{\cos\left(\frac{\pi\beta t}{T}\right)}{1-\left(\frac{2\beta t}{T}\right)^2}w\left(\frac{t}{L_pT}\right)$, where $T$ is the sampling period, $\beta$ is the roll-off factor, $\mathrm{sinc}(t)=\frac{\sin(\pi t)}{\pi t}$ is the sinc function, and $w(t)$ is a window function such that $w(t)=1$ for $\lvert t\rvert\leq1$ and $w(t)=0$ otherwise. 
Discretizing the composite channel in (\ref{composite channel}), we have\footnote{Unlike in Fig. \ref{leakage}, we do not account for the delay shift because we can always shift the taps at the end of the CIR back to the beginning.} 
\begin{equation}\label{discrete channel}
\begin{aligned}
    h_n=h(nT)=\sum_{\ell=0}^{L-1}\alpha_\ell g(nT-\tau_\ell),~~n=0,\cdots,N-1,
\end{aligned}
\end{equation}
where $N$ corresponds to the maximum delay spread of the composite channel.  The leakage effect is evident from (\ref{discrete channel}) as the energy of a path leaks to multiple taps in a pulse, making it challenging to figure out the exact delay of the physical path. 
For a WiFi OFDM system with $K$ subcarriers, the frequency response of the composite channel on subcarrier $k$ is given by
\begin{equation}
\begin{aligned}
    H_k=\sum_{n=0}^{N-1}h_ne^{-j2\pi kn/N},~~k=0,\cdots,K-1.\\
\end{aligned}
\end{equation}

In practical WiFi systems, only a subset of subcarriers are used to transmit data/pilot symbols. For example, in IEEE 802.11a, the $0$-th subcarrier is left unused due to the strong DC interference, and subcarriers indexed from 27 to 37 are also excluded to avoid interference from neighboring channels. In this paper, we estimate multipath delays from the CSI on those subcarriers available for data/pilot transmission.

\section{SBL-Based Delay Estimation}
In this section, we first leverage the knowledge of the pulse shape to formulate multipath delay estimation as a sparse recovery problem by discretizing the delay parameters into grid points, and then propose an SBL-based approach to estimate the delay parameters of the physical paths.
\subsection{Problem Formulation}
Denote $\mathbf{h}_d=[h_0,\cdots,h_{N-1}]^T$ and $\mathbf{h}_f=[H_0,\cdots,H_{K-1}]^T$. The relationship between $\mathbf{h}_f$ and $\mathbf{h}_d$ can be written as
\begin{equation}\label{Fourier}
\mathbf{h}_f=\mathbf{F}_{0:N-1}\mathbf{h}_d,
\end{equation}
where $\mathbf{F}_{0:N-1}$ is a partial discrete Fourier transform (DFT) matrix composed of the first $N$ columns of a complete DFT matrix $\mathbf{F}$ of size $K$, with the $(m,n)$-th entry of $\mathbf{F}$ given by $F_{m,n}=e^{-j2\pi mn/N}$. As mentioned in Section \ref{sigmal model}, only a portion of the frequency response is available. Therefore, the measured CSI can be expressed as
\begin{equation}\label{measured CSI}
\mathbf{y}=\mathbf{F}_{\mathcal{I},0:N-1}\mathbf{h}_d+\mathbf{n},
\end{equation}
where $\mathbf{y}$ is the measured CSI, $\mathcal{I}$ is the index set of subcarriers used for data/pilot transmission, $\mathbf{F}_{\mathcal{I},0:N-1}$ is a matrix composed of $|\mathcal{I}|$ rows of $\mathbf{F}_{0:N-1}$ corresponding to the used subcarriers, and $\mathbf{n}$ is the complex Gaussian-distributed measurement error with zero mean and variance $\sigma^2$. 

Taking into account the knowledge of the pulse shape, $\mathbf{h}_d$ can be further written as
\begin{equation}\label{pulse expansion}
\mathbf{h}_d=\mathbf{\bar{A}}\mathbf{\bm{\bar{\alpha}}},
\end{equation}
where $\mathbf{\bar{A}}\in\mathbb{R}^{N\times L}$ and its $(n,\ell)$-th entry is given by $\bar{A}_{n,\ell}=g(nT-\tau_\ell)$, and $\bm{\bar{\alpha}}=[\alpha_0,\cdots,\alpha_{L-1}]^T$. Motivated by the expression in (\ref{pulse expansion}), we can discretize the delay parameter into a set of fine-grained grid points as $\bm{\tau}=[0,\cdots,(M-1)T_g]^T$, where $T_g$ is the resolution of the grid, $M$ is the number of grid points, and $(M-1)T_g$ is the maximum potential delay spread of the physical paths. Then, we can construct a dictionary matrix $\mathbf{A}\in\mathbb{R}^{N\times M}$ with its $(n,m)$-th entry given by $A_{n,m}=g(nT-mT_g)$, and (\ref{pulse expansion}) can be approximated as
\begin{equation}\label{grid representation}
\mathbf{h}_d\approx\mathbf{A}\mathbf{\bm{\alpha}}.
\end{equation}
In (\ref{grid representation}), when the delay of a path falls on a specific grid point, the corresponding entry in $\mathbf{\bm{\alpha}}$ will be nonzero. Due to the limited number of paths in the physical environment, there are only a small fraction of nonzero entries in $\mathbf{\bm{\alpha}}$. In other words, $\mathbf{\bm{\alpha}}$ is sparse. Substituting (\ref{grid representation}) into (\ref{measured CSI}), we have
\begin{equation}\label{sparse}
\mathbf{y}\approx\mathbf{F}_{\mathcal{I},0:N-1}\mathbf{A}\mathbf{\bm{\alpha}}+\mathbf{n}=\mathbf{B}\mathbf{\bm{\alpha}}+\mathbf{n},
\end{equation}
where $\mathbf{B}\triangleq\mathbf{F}_{\mathcal{I},0:N-1}\mathbf{A}$. Next, our objective is to recover $\mathbf{\bm{\alpha}}$ from $\mathbf{y}$ given $\mathbf{B}$. After $\mathbf{\bm{\alpha}}$ is obtained, we can determine the delay parameters from the positions of its nonzero entries.

\subsection{Delay Estimation}

In this part, we estimate $\bm{\alpha}$ in (\ref{sparse}) using an SBL-based method \cite{tipping2001sparse}, which is a statistical approach for sparse signal estimation, and has been shown to be less susceptible to \textit{structural error} and \textit{convergence error} compared to other sparse recovery algorithms \cite{wipf2004sparse}.

To proceed, we assign a complex Gaussian prior distribution to $\bm{\alpha}$:
\begin{equation}\label{prior}
p(\bm{\alpha}|\bm{\gamma})=\prod_{m=0}^{M-1}\mathcal{CN}(\alpha_m;0,\gamma^{-1}_m)=\prod_{m=0}^{M-1}\frac{\gamma_m}{\pi}e^{-\gamma_m|\alpha_m|^2},
\end{equation}
where $\bm{\gamma}=[\gamma_0,\cdots,\gamma_{M-1}]^T$, $\mathcal{CN}(\mathbf{x};\bm{\mu},\bm{\Sigma})$ represents the probability density function of a complex Gaussian random vector (or a random variable in the scalar case) with mean $\bm{\mu}$ and covariance matrix $\bm{\Sigma}$. Since $\bm{\gamma}$ is not known \textit{a priori}, we adopt a Gamma hyperprior model because it is the conjugate prior to the Gaussian distribution in (\ref{prior}):
\begin{equation}\label{hyperprior variance}
    p(\bm{\gamma})=\prod_{m=0}^{M-1}\mathrm{Gamma}(\gamma_m;a,b)=\prod_{m=0}^{M-1}\frac{b^a}{\Gamma(a)}\gamma_m^{a-1}e^{-b\gamma_m},
\end{equation}
where $\mathrm{Gamma}(\gamma;a,b)$ is the probability density function of a Gamma-distributed variable $\gamma$ with shape parameter $a$ and rate parameter $b$, and $\Gamma(\cdot)$ is the Gamma function. In general, the parameters $a$ and $b$ should take small values to make the hyperprior model non-informative.
On the other hand, based on (\ref{sparse}), the likelihood function is given by
\begin{equation}
p(\mathbf{y}|\bm{\alpha},\beta)=\mathcal{CN}(\mathbf{y};\mathbf{B}\bm{\alpha},\beta^{-1}\mathbf{I})=\frac{\beta^{\lvert\mathcal{I}\rvert}}{\pi^{\lvert\mathcal{I}\rvert}}e^{-\beta\lVert\mathbf{y}-\mathbf{B}\bm{\alpha}\rVert_2^2},
\end{equation}
where $\beta=\frac{1}{\sigma^2}$, and $\mathbf{I}$ is an identity matrix. We assign another Gamma distribution to $\beta$ since it is also unknown:
\begin{equation}\label{beta}
    p(\beta)=\mathrm{Gamma}(\beta;c,d)=\frac{d^c}{\Gamma(c)}\beta^{c-1}e^{-d\beta}.
\end{equation}
The joint probability density of all the involved variables can be written as
\begin{equation}\label{joint probability}
p(\mathbf{y},\bm{\alpha},\bm{\gamma},\beta)=p(\mathbf{y}|\bm{\alpha},\beta)p(\bm{\alpha}|\bm{\gamma})p(\bm{\gamma})p(\beta).
\end{equation}

With the probabilistic model established, we need to jointly estimate the posterior distributions of $\bm{\alpha}$, $\bm{\gamma}$ and $\beta$ given $\mathbf{y}$ from (\ref{joint probability}). The sparsity pattern of $\bm{\alpha} $ can then be reflected from the estimate of $\bm{\gamma}$. Specifically, when $\gamma_m$ is large, $\alpha_m$ is forced to be around zero; when $\gamma_m$ is small, $\alpha_m$ is likely to be large. Therefore, we can find the delay of a path from the position where $\gamma_m$ takes a small value. It is worth mentioning that the Gaussian prior model in (\ref{prior}) and its hyperprior in (\ref{hyperprior variance}) are used only to enhance the sparsity, and the actual distribution of the multipath amplitudes is not necessarily identical to the assumed prior.

Since the posterior distribution of the involved variables is difficult to compute directly, we use the variational inference approach \cite{tzikas2008variational} to approximate it. The variational inference assumes that the joint posterior distribution can be factorized~as
\begin{equation}
p(\bm{\alpha},\bm{\gamma},\beta|\mathbf{y})\approx q(\bm{\alpha})q(\bm{\gamma})q(\beta).
\end{equation}
In other words, the variables are mutually independent. By maximizing a lower bound of the likelihood function, the logarithm of the approximated posterior distribution of each variable can be expressed as\cite{tzikas2008variational}
\begin{equation}\label{posterior}
\ln q(z_i)=\int\ln p(\mathbf{y},\bm{\alpha},\bm{\gamma},\beta)\prod_{j\neq i}q(z_j)\mathrm{d}z_j+\mathrm{const.},
\end{equation}
where $z_i$ represents one of the variables $\bm{\alpha}$, $\bm{\gamma}$ or $\beta$, and ``$\mathrm{const.}$" is a constant irrelevant to $z_i$. The expression in (\ref{posterior}) implies an iterative implementation of the variational inference algorithm, because the posterior distribution of one variable depends on the expressions of all the others. Given that explicit expressions of the posterior distributions in the real-valued case have been provided in \cite{tzikas2008variational}, the results for the complex channel model can be obtained by making modifications to these expressions, as presented below:

\begin{itemize}
\item The posterior distribution of $\bm{\alpha}$ is complex Gaussian with the covariance matrix $\bm{\Sigma}$ and mean vector $\bm{\mu}$ given by
\begin{equation}\label{posterior alpha}
\bm{\Sigma}=(\langle\beta\rangle\mathbf{B}^H\mathbf{B}+\mathbf{D})^{-1},~~
\bm{\mu}=\langle\beta\rangle\bm{\Sigma}\mathbf{B}^H\mathbf{y},
\end{equation}
where $\langle\cdot\rangle$ is the expectation operation with respect to the posterior distribution, and $\mathbf{D}=\mathrm{diag}\{\langle\bm{\gamma}\rangle\}$, which is a diagonal matrix with $\langle\bm{\gamma}\rangle=[\langle\gamma_0\rangle,\cdots,\langle\gamma_{M-1}\rangle]^T$ on its diagonal. 

\item The posterior distribution of $\gamma_m~(m=0,\cdots,M-1)$ is a Gamma distribution with the shape parameter $\tilde{a}_m$ and rate parameter $\tilde{b}_m$ expressed as
\begin{equation}\label{posterior gamma}
\tilde{a}_m=a+1,~~
\tilde{b}_m=b+\langle\lvert\alpha_m\rvert^2\rangle,
\end{equation}
where $\langle\lvert\alpha_m\rvert^2\rangle=\lvert\mu_m\rvert^2+\Sigma_{m,m}$, with $\mu_m$ and $\Sigma_{m,m}$ defined as the $m$-th entry of $\bm{\mu}$ and the $m$-th diagonal entry of $\bm{\Sigma}$, respectively. The posterior expectation of $\gamma_m$ is then given by $\langle\gamma_m\rangle=\tilde{a}_m/\tilde{b}_m$.
\item The posterior distribution of $\beta$ is also a Gamma distribution with the shape parameter $\tilde{c}$ and rate parameter $\tilde{d}$ expressed as
\begin{equation}\label{posterior beta}
\tilde{c}=c+\lvert\mathcal{I}\rvert,~~
\tilde{d}=d+\langle\lVert\mathbf{y}-\mathbf{B}\bm{\alpha}\rVert^2_2\rangle,
\end{equation}
where $\langle\lVert\mathbf{y}-\mathbf{B}\bm{\alpha}\rVert^2_2\rangle=\lVert\mathbf{y}-\mathbf{B}\bm{\mu}\rVert^2_2+\mathrm{tr}(\mathbf{B}\bm{\Sigma}\mathbf{B}^H)$ with $\mathrm{tr}(\cdot)$ denoting the trace of a matrix. Similar to $\gamma_m$, the posterior expectation of $\beta$ is given by $\langle\beta\rangle=\tilde{c}/\tilde{d}$.
\end{itemize}

The computational complexity of the SBL-based algorithm primarily arises from the matrix inverse in $\bm{\Sigma}$, which is of order $\mathcal{O}(M^3)$ in each iteration. We can use the Woodbury inversion identity to reduce the computational burden\cite{tipping2001sparse}:
\begin{equation}\label{woodbury}
\begin{aligned}
&(\langle\beta\rangle\mathbf{B}^H\mathbf{B}+\mathbf{D})^{-1}\\
&=\mathbf{D}^{-1}-\langle\beta\rangle\mathbf{D}^{-1}\mathbf{B}^H(\mathbf{I}+\langle\beta\rangle\mathbf{B}\mathbf{D}^{-1}\mathbf{B}^H)^{-1}\mathbf{B}\mathbf{D}^{-1}.
\end{aligned}
\end{equation}
Consequently, the complexity of the matrix inversion is reduced to $\mathcal{O}(|\mathcal{I}|^3)$. Since $M\gg|\mathcal{I}|$ in this paper, the matrix multiplication will then dominate the complexity, which is $\mathcal{O}(M^2|\mathcal{I}|)$. Furthermore, as the algorithm progresses, we can prune some basis vectors whose corresponding values of $\langle\gamma_m\rangle$ are significantly large. Specifically, we remove the $m$-th column of $\mathbf{B}$ and the $m$-th entry of $\bm{\alpha}$ if $\langle\gamma_m\rangle$ is larger than a threshold.

After the algorithm converges, the grid points corresponding to the remaining basis vectors can be used as the estimate of delay parameters. If the number of the remaining basis vectors exceeds a predetermined value $L_{\mathrm{max}}$, we select a subset of $L_{\mathrm{max}}$ delay estimates with the smallest values of $\langle\gamma_m\rangle$. Then, we further choose the delay estimates that are separated by at least $\Delta\tau$ from each other. 
All the above procedures are summarized in Algorithm \ref{alg1}.\footnote{The stopping criterion for the while loop is defined by either reaching the maximum number of iterations or attaining a relative change in the estimated path amplitudes below a threshold, which will be given in Section \ref{simulation}.} If the complex amplitudes of the paths are of interest as well,  we can fix the delay estimates and reexecute the algorithm until convergence. The posterior mean of $\bm{\alpha}$ can then be used as an accurate estimate of the path amplitudes. As a consequence, a ``cleaned" version of the CSI data can also be reconstructed.

\begin{algorithm}[h]
\small
\caption{SBL-Based Multipath Delay Estimation}
\label{alg1}
\begin{algorithmic}
\STATE \textbf{Input:} 
Measured CSI $\mathbf{y}$, hyperparameters $a,b,c,d$, the delay grid resolution $T_g$, the threshold for deleting a basis vector $\eta$, and the spacing for delay selection $\Delta\tau$.
\STATE \textbf{Output:} The delay estimates $\mathcal{T}$.
\STATE \textbf{Initialization:}  The delay estimates $\mathcal{T}=\{0,\cdots,(M-1)T_g\}\triangleq\{\hat{\tau}_0,\cdots,\hat{\tau}_{M-1}\}$, the matrix $\mathbf{B}$, $\langle\gamma_m\rangle=a/b$ for all $m$, and $\langle\beta\rangle=c/d$.
\WHILE{$stopping~criterion~not~met$}
\STATE Compute the posterior distribution of $\bm{\alpha}$ using (\ref{posterior alpha}).
\STATE Compute the posterior distribution of $\bm{\gamma}$ using (\ref{posterior gamma}).
\STATE Compute the posterior distribution of $\beta$ using (\ref{posterior beta}).
\STATE Delete $\{\hat{\tau}_m:\langle\gamma_m\rangle>\eta\min\{\langle\gamma_{m'}\rangle\}_{m'=0}^{\lvert\mathcal{T}\rvert-1}\}$ from $\mathcal{T}$.
\STATE Reconstruct $\mathbf{B}$ and $\bm{\gamma}$ using $\mathcal{T}$.
\ENDWHILE
\IF{$\lvert\mathcal{T}\rvert>L_{\mathrm{max}}$}
\STATE Retain only $L_{\mathrm{max}}$ elements in $\mathcal{T}$ with small values of $\langle\gamma_m\rangle$.
\ENDIF
\STATE Order $\{\langle\gamma_m\rangle\}_{m=0}^{\lvert\mathcal{T}\rvert-1}$ such that $\langle\gamma_{m_0}\rangle\leq\cdots\leq\langle\gamma_{m_{\lvert\mathcal{T}\rvert-1}}\rangle$.
\FOR{$i=1:\lvert\mathcal{T}\rvert-1$}
\IF{$\lvert\hat{\tau}_{m_i}-\hat{\tau}_{m_0}\rvert\leq\Delta\tau~\OR\cdots~\OR~\lvert\hat{\tau}_{m_i}-\hat{\tau}_{m_{i-1}}\rvert\leq\Delta\tau$}
\STATE Delete $\hat{\tau}_{m_i}$ from $\mathcal{T}$.
\ENDIF
\ENDFOR
\end{algorithmic}
\end{algorithm}

\section{Simulations}\label{simulation}
In this section, the performance of the proposed multipath delay estimation algorithm is evaluated through simulations. We use the MATLAB WLAN Toolbox to generate multipath channels. The system bandwidth is $20\,\mathrm{MHz}$, corresponding to a sampling period $T=50\,\mathrm{ns}$. Out of a total number of $K=64$ subcarriers, 52 are used for data/pilot transmission with indices from 1 to 26 and from 38 to 63. The length of the cyclic prefix is set to $32$, which is the same as the length of the delay-domain composite channel. We use a raised-cosine pulse function with the roll-off factor $\rho=0.05$, truncated to a nonzero duration of $16T$. In other words, $L_p=8$. The number of paths is $3$ with delay parameters being set as $24\,\mathrm{ns}$, $65\,\mathrm{ns}$ and $103\,\mathrm{ns}$. Note that the delay difference between adjacent paths is smaller than the sampling period. The signal-to-noise ratio (SNR) is defined as the ratio between the power of the first path to the variance of the measurement error $\sigma^2$, and the power of each subsequent path is 5 dB smaller than that of its previous one. For any unspecified parameters, we use the default settings in MATLAB. In the proposed delay estimation algorithm, the grid resolution is set to $T_g=1\,\mathrm{ns}$. Hence, the number of grid points is $M=850$. The hyperparameters in the Gamma distribution are given by $a=b=c=d=1\times 10^{-6}$, and we initialize $\langle\alpha_m\rangle=a/b$ and $\langle\beta\rangle=c/d$. A threshold of $\eta=1\times10^5$ is chosen for deleting a basis vector.  After the iterations, we set $L_\mathrm{max}=10$ and $\Delta\tau=5~\mathrm{ns}$ for further delay selection. 

Two benchmarks are considered in this paper: the space-alternating generalized expectation-maximization (SAGE) algorithm \cite{fleury1999channel} and the OMP-based method \cite{scaglione2011compressed}. The SAGE algorithm is an iterative method to estimate multipath parameters and has been used in \cite{zhang2022toward,qian2018widar2} for WiFi sensing. However, it neglects the channel leakage effect. For the OMP-based method, it takes into account the pulse shaping function and adopts a greedy approach to find the nonzero positions in $\bm{\alpha}$. In theory, all three algorithms can overcome the low-precision limitation resulting from the restricted sampling rate of WiFi systems. 
We input the ground truth of the number of paths into the two benchmarks, resulting in an upper bound for performance evaluation. For the SAGE and our proposed algorithms, the iteration is terminated when either the relative change of the estimated path amplitudes is less than $1\times10^{-4}$ or the number of iterations reaches 1000, while the number of iterations of OMP is fixed to $L$. All the results presented below are obtained by averaging over 2000 simulations. The computational complexity of the SAGE, the OMP, and the proposed algorithms are $\mathcal{O}(tLM|\mathcal{I}|)$, $\mathcal{O}(LM|\mathcal{I}|)$ and $\mathcal{O}(tM^2|\mathcal{I}|)$, respectively, where $t$ is the number of iterations actually used. It should be noted that we use an exhaustive search method to find the optimal delay estimate in the maximization step of SAGE, the code of which is released by the authors of \cite{qian2018widar2}. Despite a higher complexity, we will show that the proposed algorithm can achieve markedly better performance than the two benchmarks.

In Table \ref{tab1}, we show the probability of correct estimation and the mean absolute error (MAE) of the number of paths of the proposed algorithm. As we can see, when the SNR is high, our algorithm can correctly determine the number of paths with a high probability and a low MAE.
When the SNR is not large enough, the power of some paths with large delays could be even lower than the noise variance due to the exponential decay. As a result, they cannot be detected and the estimated number of paths is typically smaller than the ground truth. In sensing applications, such weak paths are usually discarded even if detected, because they cannot provide reliable information about a target.

\begin{table}
\caption{Probability of correct estimation of the number of paths}
\label{tab1}
\centering
\footnotesize
\begin{tabular}{|c|c|c|c|c|c|c|c|}
\hline
\text{SNR}&20&25&30&35&40 \\
\hline
\text{Probability}&0.7475&0.8630&0.8990&0.9000&0.9145\\
\hline
\text{MAE}&0.2535&0.1370&0.1010&0.1020&0.0875\\
\hline
\end{tabular}
\vspace{-1em}
\end{table}

\begin{figure}[t]
\centering
\includegraphics[width=0.5\columnwidth]{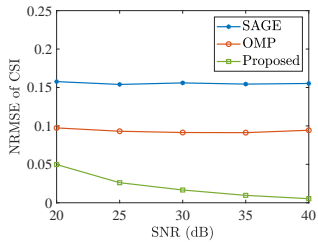}
\caption{Comparison of the NRMSE of CSI.} 
\label{CSI}
\vspace{-1.5em}
\end{figure}

\begin{figure}[t]
\centering
\setlength{\textfloatsep}{1pt plus 1pt minus 1pt}
\setlength{\floatsep}{1pt plus 1pt minus 1pt}
\includegraphics[width=0.5\columnwidth]{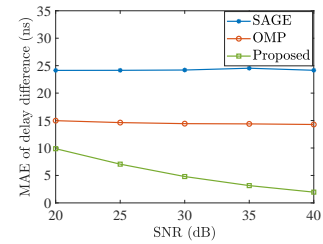}
\caption{Comparison of the MAE of $\tau_0-\tau_1$.}
\label{delay}
\vspace{-1.5em}
\end{figure}

Fig. \ref{CSI} and Fig. \ref{delay} compare the normalized root mean square error (NRMSE) of the CSI reconstruction and the MAE of the delay parameter estimation, respectively. Since the absolute delay information is lost after the packet detection (see Fig. \ref{leakage}), we compute the estimation error of the delay difference between the first and the second paths. We can see that the proposed SBL-based algorithm outperforms the two benchmarks under both metrics. In SAGE, the effect of pulse shaping is neglected, and fake paths could be estimated, resulting in the worst performance. Although the OMP-based method uses the knowledge of the pulse shape, its greedy strategy can produce an inaccurate delay estimate, especially when basis vectors are highly correlated in our sparse model, and the estimation error cannot be corrected in subsequent iterations, leading to an inferior performance compared to the proposed method. The above factors also prevent the performance improvements of the two benchmarks as the SNR increases. In contrast, the proposed algorithm not only takes into account the pulse shaping function, but also adaptively updates the weights of the delay grid points, and thus achieves the best performance.

\section{Conclusions}
This paper investigated the multipath delay estimation problem in leaked WiFi channels, in which a physical path may produce multiple taps in the delay domain. We leveraged the knowledge of the pulse shape and formulated a sparse recovery problem using a dictionary matrix composed of discretized pulses shifted by different delays. We used the SBL method to solve this problem, and the path delay parameters can be automatically identified from the nonzero positions of the recovered sparse vector. Simulation results have shown that our proposed algorithm can accurately estimate the number of paths and outperforms two counterparts in terms of the RNMSE of CSI reconstruction and the MAE of delay estimation.
\bibliography{document}

\begin{thebibliography}{10}
\providecommand{\url}[1]{#1}
\csname url@samestyle\endcsname
\providecommand{\newblock}{\relax}
\providecommand{\bibinfo}[2]{#2}
\providecommand{\BIBentrySTDinterwordspacing}{\spaceskip=0pt\relax}
\providecommand{\BIBentryALTinterwordstretchfactor}{4}
\providecommand{\BIBentryALTinterwordspacing}{\spaceskip=\fontdimen2\font plus
\BIBentryALTinterwordstretchfactor\fontdimen3\font minus
  \fontdimen4\font\relax}
\providecommand{\BIBforeignlanguage}[2]{{%
\expandafter\ifx\csname l@#1\endcsname\relax
\typeout{** WARNING: IEEEtran.bst: No hyphenation pattern has been}%
\typeout{** loaded for the language `#1'. Using the pattern for}%
\typeout{** the default language instead.}%
\else
\language=\csname l@#1\endcsname
\fi
#2}}
\providecommand{\BIBdecl}{\relax}
\BIBdecl

\bibitem{ma2019wifi}
Y.~Ma, G.~Zhou, and S.~Wang, ``Wi{F}i sensing with channel state information: A
  survey,'' \emph{ACM Computing Surveys (CSUR)}, vol.~52, no.~3, pp. 1--36,
  2019.

\bibitem{tan2022commodity}
S.~Tan, Y.~Ren, J.~Yang, and Y.~Chen, ``Commodity {W}i{F}i sensing in ten
  years: Status, challenges, and opportunities,'' \emph{IEEE Internet of Things
  Journal}, vol.~9, no.~18, pp. 17\,832--17\,843, 2022.

\bibitem{halperin2011tool}
D.~Halperin, W.~Hu, A.~Sheth, and D.~Wetherall, ``Tool release: Gathering
  802.11n traces with channel state information,'' \emph{ACM SIGCOMM Computer
  Communication Review}, vol.~41, no.~1, pp. 53--53, 2011.

\bibitem{xie2015precise}
Y.~Xie, Z.~Li, and M.~Li, ``Precise power delay profiling with commodity
  {W}i{F}i,'' in \emph{Proceedings of the 21st Annual International Conference
  on Mobile Computing and Networking}, 2015, pp. 53--64.

\bibitem{taubock2010compressive}
G.~Taubock, F.~Hlawatsch, D.~Eiwen, and H.~Rauhut, ``Compressive estimation of
  doubly selective channels in multicarrier systems: Leakage effects and
  sparsity-enhancing processing,'' \emph{IEEE Journal of Selected Topics in
  Signal Processing}, vol.~4, no.~2, pp. 255--271, 2010.

\bibitem{zhang2022toward}
X.~Zhang, L.~Chen, M.~Feng, and T.~Jiang, ``Toward reliable non-line-of-sight
  localization using multipath reflections,'' \emph{Proceedings of the ACM on
  Interactive, Mobile, Wearable and Ubiquitous Technologies}, vol.~6, no.~1,
  pp. 1--25, 2022.

\bibitem{li2022superresolution}
Z.~Li, A.~Nimr, P.~Schulz, and G.~Fettweis, ``Superresolution wireless
  multipath channel path delay estimation for {CIR}-based localization,'' in
  \emph{2022 IEEE Wireless Communications and Networking Conference
  (WCNC)}.\hskip 1em plus 0.5em minus 0.4em\relax IEEE, 2022, pp. 1940--1945.

\bibitem{li2021improved}
J.~Li and U.~Mitra, ``Improved atomic norm based time-varying multipath channel
  estimation,'' \emph{IEEE Transactions on Communications}, vol.~69, no.~9, pp.
  6225--6235, 2021.

\bibitem{tipping2001sparse}
M.~E. Tipping, ``Sparse {B}ayesian learning and the relevance vector machine,''
  \emph{Journal of Machine Learning Research}, vol.~1, no. Jun, pp. 211--244,
  2001.

\bibitem{stoica2004model}
P.~Stoica and Y.~Selen, ``Model-order selection: a review of information
  criterion rules,'' \emph{IEEE Signal Processing Magazine}, vol.~21, no.~4,
  pp. 36--47, 2004.

\bibitem{scaglione2011compressed}
A.~Scaglione and X.~Li, ``Compressed channel sensing: Is the restricted
  isometry property the right metric?'' in \emph{2011 17th International
  Conference on Digital Signal Processing (DSP)}.\hskip 1em plus 0.5em minus
  0.4em\relax IEEE, 2011, pp. 1--8.

\bibitem{wipf2004sparse}
D.~P. Wipf and B.~D. Rao, ``Sparse {B}ayesian learning for basis selection,''
  \emph{IEEE Transactions on Signal Processing}, vol.~52, no.~8, pp.
  2153--2164, 2004.

\bibitem{tzikas2008variational}
D.~G. Tzikas, A.~C. Likas, and N.~P. Galatsanos, ``The variational
  approximation for {B}ayesian inference,'' \emph{IEEE Signal Processing
  Magazine}, vol.~25, no.~6, pp. 131--146, 2008.

\bibitem{fleury1999channel}
B.~H. Fleury, M.~Tschudin, R.~Heddergott, D.~Dahlhaus, and K.~I. Pedersen,
  ``Channel parameter estimation in mobile radio environments using the {SAGE}
  algorithm,'' \emph{IEEE Journal on Selected Areas in Communications},
  vol.~17, no.~3, pp. 434--450, 1999.

\bibitem{qian2018widar2}
K.~Qian, C.~Wu, Y.~Zhang, G.~Zhang, Z.~Yang, and Y.~Liu, ``Widar2. 0: Passive
  human tracking with a single {W}i-{F}i link,'' in \emph{Proceedings of the
  16th Annual International Conference on Mobile Systems, Applications, and
  Services}, 2018, pp. 350--361.

\end{thebibliography}
\bibliographystyle{IEEEtran}
\end{document}